\documentclass[%
 reprint,
 amsmath,amssymb,
 aps,superscriptaddress
]{revtex4-2}

\usepackage{graphicx}
\usepackage{dcolumn}
\usepackage{bm}
\usepackage[normalem]{ulem}
\usepackage{xcolor}
\usepackage{mathrsfs}
\usepackage{inputenc}
\usepackage[slantedGreek]{mathpazo}
\usepackage{hyperref}
\begin{document}
\newcommand{\rb}[1]{\textcolor{red}{\it#1}}
\newcommand{\rbout}[1]{\textcolor{red}{\sout{#1}}}

\preprint{APS/123-QED}

\title{Effective electric field associated with the electric dipole moment of the electron for TlF$^+$}

\author{R. Bala}
\email{balar180@gmail.com}
\affiliation{Centre for Quantum Engineering, Research and Education, TCG CREST, Salt Lake, Kolkata 700091, India}
\author{V. S. Prasannaa}%
 \email{srinivasa.prasannaa@tcgcrest.org}
\affiliation{Centre for Quantum Engineering, Research and Education, TCG CREST, Salt Lake, Kolkata 700091, India}%
\author{M. Abe}%
\affiliation{Quantum Chemistry Research Laboratory, Department of Chemistry, School of Science, Hiroshima University, Higashi-hiroshima, Hiroshima 739-8511, Japan}%
\author{B. P. Das}%
\affiliation{Centre for Quantum Engineering, Research and Education, TCG CREST, Salt Lake, Kolkata 700091, India}%

\begin{abstract}

In this article, we have employed relativistic many-body theory to theoretically assess the suitability of TlF$^+$ molecular ion in its ground state for electron electric dipole moment searches. To that end, we have computed values of the effective electric field as well as the molecular permanent electric dipole moment using both configuration interaction and coupled cluster methods with high quality basis sets, followed by an analysis on the role of electron correlation in the considered properties. We find that TlF$^+$ has a large value of effective electric field of about 163 GV/cm, which is about one and a half times larger than the HgF and HgH  molecules, which are known to have the largest effective electric fields among non-superheavy systems.  

\begin{description}

\item[Keywords]
Electric dipole moment of the electron, effective electric field, molecular electric dipole moment, relativistic coupled cluster method, Kramers restricted configuration interaction. 
\end{description}
\end{abstract}

\maketitle

\section{\label{sec:level1} Introduction}

The past decade has witnessed proposing and/or  commencing of several high-precision experiments in the non-accelerator particle physics sector \cite{Jaeckel_2020, Wu_2022, Safronova_2018, Verma_2020, Klapdor_2002}, where one employs atoms or molecules to search for new physics. In particular, there has been a sharp rise in experimental and theoretical efforts to search for the elusive and yet-undetected parity- ($\mathcal P$) and time reversal- ($\mathcal T$) violating electric dipole moment (EDM) of the electron using heavy polar molecules \cite{Vutha_2018, Loh_2013, Aggrawal_2018, Campbell_2013, Fitch_2020, Tarbutt_2009}. Electron EDM (eEDM, denoted by $d_e$) experiments using molecules play a crucial role in the understanding of physics beyond the standard model of particle physics~\cite{Fukuyama_2012, Khriplovich_1997} and also shed light on the baryon asymmetry in the universe \cite{Fuyoto_BAU_2016, Kazarian_1992}, by accessing energy scales that are unattainable by current-day accelerators \cite{Reece}. There are currently several ongoing eEDM experiments, which employ ThO, YbF, HfF$^+$, BaF, or YbOH \cite{Hudson_2011, Baron_2017,Cairncross_2017, Aggrawal_2018, Timothy_2018}, with the ThO experiment setting the best upper bound on eEDM ($\sim$ 10$^{-29}$ e-cm)~\cite{Baron_2014}. Several theoretical works have proposed numerous molecular candidates, including HgX (X = F, Cl, Br, and I), HgA (A = Li, Na, and K), YbZ (Z = Rb, Cs, Sr$^+$, and Ba$^+$) RaF, TaN, YbOH, BaOH, RaH, WC, TaO$^+$,  etc.~\cite{Meyer_WC_2009, Mitra_2021, kudashov_2014, Prasannaa_2015, Skripnikov_TaN_2015, Shafer_2006, Meyer_YbX_2009, Sunaga_monoF_2018, Fazil_RaH_2019, Denis_YbOH_2019, Sunaga_HgA_2019, Fleig_TaO+_2017}. \\

A very desirable feature for proposing new molecular candidates for future eEDM searches is a large effective electric field of a candidate molecule. This is because in the expression for the figure of merit for the projected statistical sensitivity of a proposed eEDM experiment, the effective electric field, $E_{eff}$, occurs $\emph{outside}$ the square root, as opposed to other factors such as the total number of molecules per second or the coherence time, which occur inside the square root~\cite{Khriplovich_1997}. Hence, $E_{eff}$ plays a vital role in deciding the suitability of a candidate molecule for EDM searches. It is worth noting that the molecular electric dipole moment (abbreviated hereafter as PDM in this work) also plays an important role in the projected sensitivity of an EDM experiment via the polarizing factor~\cite{Prasannaa_2015}. \\

In this work, we investigate the effective electric field and the PDM of the TlF$^+$ molecule in its ground state by using relativistic many-body theory. In particular, we employ Kramers restricted configuration interaction (KRCI) method and the relativistic coupled cluster (RCC) method for this purpose. It is worth adding at this point that another work independently calculates the effective electric field of TlF$^+$ molecule~\cite{Jungblut_2022}. 
\section{Theory}
In this section, we very briefly discuss the relevant theoretical background for our work, namely the properties of interest to us (effective electric field and PDM) as well as the relativistic many-body theories that we employ in this work (KRCI and RCC methods). Further details can be found in Refs.~\cite{Nayak_KRCI_2013, Prasannaa_2015, DIRAC_code2020, Sunaga_HgA_2019}.
\subsection{$\mathcal P~\&~\mathcal T$ -odd interaction constant relevant to eEDM}
In a single valence molecule, $E_{eff}$ arises from relativistic interactions of the EDM of the unpaired electron with the electric fields created due to all other charged particles in that molecular system. The expectation value of the eEDM operator, $H_{EDM}$, is given by \cite{Das_1989, Commins, Prasannaa_2015, Abe_YbF_2014}
\begin{eqnarray}
 \Delta U=\bigg\langle\sum\limits_{k = 1}^{n_e}{H}_{EDM}(k)\bigg\rangle_{\Psi} &=& -d_e\bigg\langle\sum\limits_{k=1}^{n_e} \gamma_k^0 \Sigma_k \cdot\varepsilon_k\bigg\rangle_{\Psi} \nonumber\\
&=&\frac{2icd_e}{e\hbar}\bigg\langle\sum\limits_{k = 1}^{n_e}\gamma_k^0 \gamma_k^5 p_k^2\bigg\rangle_{\Psi}, 
\end{eqnarray}
where $n_e$ is the number of electrons in the molecule; $\gamma^0$ and $\gamma^5$ are $4$\,-\,component Dirac matrices; $\varepsilon_k$ is the electric field at the position of $k^{th}$ electron; $p_k$ is the momentum operator; and $|\Psi\rangle$ is the wave function determined from relativistic many\,-\,body theory. Finally, the effective electric field experienced by the unpaired electron in the molecular system is defined as 
\begin{eqnarray}{\label{Eeff}}
 E_{eff}\,=\,W_d\Omega, 
\end{eqnarray}
Here $W_d\,(=\,({2ic}/{\Omega\,e\hbar})\langle\gamma^0\,\gamma^5\,p^2\rangle_{\Psi}$) is the associated $\mathcal P$ \& $\mathcal T$\,-odd interaction constant and $\Omega\,(=1/2)$ is the \textit{z}\,-\,component of the total angular momentum of the ground state of the molecular system. The intrinsic value of eEDM is inferred from the theoretically estimated $E_{eff}$ in conjunction with the experimentally measured shift in energy ($\Delta U$) for the electronic state of a molecule using equation $\Delta U\,=-d_e\,E_{eff}$.
\subsection{Permanent electric dipole moment}
The total PDM of a heteronuclear diatomic molecule is given by~\cite{Sunaga_2016}
\begin{eqnarray}{\label{DM}}
\mu&=&\bigg\langle\sum_{A=1}^{N_A}eZ_AR_A\,-\,\sum_{k=1}^{n_e}er_k\bigg\rangle_{\Psi} \nonumber\\
&=&\mu_{nuc}\,-\,\mu_{e},
\end{eqnarray}
where $R_A$ and $r_k$ represent the position of $A^{th}$ nucleus and $k^{th}$ electron, respectively, with respect to the origin of the chosen co-ordinate system. $N_A$ refers to the number of nuclei. Note that since we employ the Born-Oppenheimer approximation, $R_A$ is fixed, and is the bond length of the molecule when one of the atoms is chosen as the origin. $\mu_{nuc}$ ($\mu_{e}$) represents the nuclear (electronic) contribution to the total PDM of a molecular system. We note that for a charged system like TlF$^+$, the PDM depends on the choice of origin. 
\subsection{Relativistic quantum many-body theory}
\subsubsection{Kramers restricted configuration interaction}
The quantum many-body wave function $\arrowvert \Psi \rangle$, which is required to calculate the molecular properties discussed above, is obtained using relativistic KRCI approach or the relativistic  coupled cluster (RCC) method. The configuration interaction (CI) wave function is given as a linear combination of determinantal functions as~\cite{Szabo}
\begin{eqnarray} \label{CI}
\arrowvert \Psi_{CI}\rangle &=& C_0 \arrowvert \Phi_0 \rangle + \sum_{ia}C_i^aa_a^\dag a_i\arrowvert \Phi_0\rangle\\ \nonumber
&+&\sum_{ij,ab}C_{ij}^{ab}a_a^{\dag}a_b^{\dag}a_ja_i\arrowvert \Phi_0\rangle + \dots, 
\end{eqnarray}
where $|\Phi_0\rangle$ is the single\,-\,determinant Dirac\,-\,Fock (DF) wave function, which is chosen as the reference wave function for our correlation calculations. The summation index $i, j,\,\dots$ are used to define holes and $a, b,\dots$ for particles. When the operator $a_a^{\dag}a_i$ acts on the reference state, the hole in spin-orbital $i$ gets annihilated and is accompanied by the creation of a particle in spin-orbital $a$. Thus, the resultant determinant is a singly excited Slater determinant, $|\Phi_i^a\rangle$ ($=a_a^\dag a_i|\Phi_0\rangle$), and the corresponding excitation amplitude in Eq.~(\ref{CI}) is $C_i^a$. Similarly, $C_{ij}^{ab}$ is the excitation amplitude for double excitation. For relativistic calculations, the wave function can be expanded in terms of $\mathcal P$ and $\bar Q$ strings of $j$ four-component spinors ${\{\Phi_i\}}$ and
$N-j$ Kramers time-reversal partners ${\{\Phi_{\bar{i}}\}}$, respectively as~\cite{Nayak_KRCI_2013, Knecht_KRCI_2010}\\
\begin{eqnarray}
|\Psi_K\rangle\,=\,\sum_I^{dimF(M, N)}C_{KI}\,|(P\bar Q)_I\rangle, 
\end{eqnarray}
where $C_{KI}$ are the expansion coefficients and $F(M, N)$ represent the dimension of truncated N-particle Fock-space sector over M molecular four-spinors. The Kramer partners $\{\Phi_i, \Phi_{\bar{i}}\}$ are related via the equations as $\hat{K} \Phi_i = \Phi_{\bar{i}}$ and $\hat{K}{\Phi_{\bar{i}}} = -\Phi_i$. The Slater determinants $|(P\bar Q)_I\rangle$ can be written in terms of strings of creation operators in second quantization as 
\begin{eqnarray}
|(P\bar Q)\rangle=P^{\dag}\bar Q^{\dag}|\Phi_0\rangle, 
\end{eqnarray}
with $P^\dag \arrowvert \Phi_0\rangle=a_{P_1}^\dag a_{P_2}^\dag \dots a_{P_j}^\dag \arrowvert \Phi_0\rangle$ and $\bar Q^\dag \arrowvert \Phi_0 \rangle = a_{\bar Q_1}^\dag {a_{\bar{Q_2}}}^\dag \dots a_{\bar{Q}_{N-j}}^\dag \arrowvert \Phi_0\rangle$. 
\subsubsection{Relativistic coupled cluster method}
The RCC wave function is defined as~\cite{Bartlett_2007, Bishop_1991, Eliav_1994}
\begin{eqnarray}{\label{psi}}
|\Psi_{RCC}\rangle\,=\,e^{T}|\Phi_0 \rangle. 
\end{eqnarray}
 In eq.~(\ref{psi}), T is the cluster operator and it can be written as 
\begin{eqnarray}
T=T_1+T_2+\cdots+T_{n_e}, 
\end{eqnarray}
where $T_1$, $T_2$, $\cdots$ represent the single, double, $\cdots$ excitation operators. In this work, we have considered single (S) and double (D) excitations (the RCCSD method) in $T$. The single and double excitation operators can be defined as
\begin{eqnarray}
T_1\,=\,\sum_{ia}t_i^a\,a_a^{\dag}a_i
\end{eqnarray}
and 
\begin{eqnarray}
T_2\,=\,\sum_{\substack{i\,\neq\,j\\ a\neq b}}t_{ij}^{ab}\,a_a^{\dag}a_b^{\dag}a_ja_i,
\end{eqnarray}
respectively, where $t_i^a$ and $t_{ij}^{ab}$ are the cluster amplitudes. Note that in the relativistic coupled cluster method, we employ the Dirac-Coulomb Hamiltonian and the single particle spin-orbitals are four-component spinors. $\arrowvert \Phi_0 \rangle$ is now the single reference wave function that is built out of these single particle spin-orbitals. 
\section{\label{sec:level3}Details of our Calculations}
To carry out the DF and RCCSD calculations, we have employed the modified UTChem~\cite{Abe_utchem_2004, Yanai_utchem_2002, Yanai_utchem_2001} and the  DIRAC08~\cite{DIRAC08} programs in tandem. The property integrals were computed using UTChem. The KRCISD calculations have been performed using DIRAC22~\cite{DIRAC22} software package. We have used Dirac-Coulomb Hamiltonian for all the  calculations reported in the current work. The Gaussian charge distribution for the nucleus is used throughout. We have utilized uncontracted correlation consistent polarized valence $\mathscr{N}$ zeta (cc-pV$\mathscr{N}$Z, $\mathscr{N}$= D, T and Q) basis sets for fluorine~\cite{Dunning_1989} and Dyall basis sets of similar quality (dyall.vnz, n= 2, 3 and 4) for thallium atom~\cite{Dyall_2006}. We will hereafter refer to the choice of $\mathscr{N}=n=2$ as simply the DZ basis, whereas choices 3 and 4 for $\mathscr{N}$ and $n$ would be referred to as TZ and QZ bases, respectively. The details of the number of basis functions in each basis set are given in Table~\ref{tab:table1}. The calculated values of equilibrium bond length for TlF$^+$ reported in literature are 2.029 \AA~\cite{Wael_2021} and 2.050 \AA~\cite{Schroder_2005}. However, since we were plagued with convergence issues at the DF level of theory in DIRAC22 as well as in UTChem, we carried out and report all our computations of both $E_{eff}$ and PDM at an offset bond length of 1.95 \AA,~throughout this work. To estimate the error in our results from our choice of an offset value of bond length, we performed RCCSD calculations at 1.80, 1.85, 1.9, and 1.95 \AA, and interpolated our values to 2.029 \AA. We have chosen offset values that are lower than 2.029 \AA,~ since we faced convergence issues all the way till 2.20 \AA\,in the other direction. We find that the error due to choice of an offset bond length (1.95 \AA) is about 10 percent for both $E_{eff}$ and the PDM.\\ 
\begin{table}[htbp]
\caption{\label{tab:table1}
Basis sets used in the present work. DZ, TZ, and QZ stand for double zeta, triple zeta, and quadruple zeta respectively, denoting the quality of a basis set. }
\begin{ruledtabular}
\begin{tabular}{llll}
 Basis set & Tl & F\\
\hline
DZ & 24s,\,20p,\,14d,\,8f & 9s,\,4p,\,1d\\
TZ & 30s,\,26p,\,17d,\,11f & 10s,\,5p,\,2d,\,1f\\
QZ & 34s,\,31p,\,21d,\,14f,\,1g & 12s,\,6p,\,3d,\,2f,\,1g\\
\end{tabular}
\end{ruledtabular}
\end{table}

To carry out KRCI calculations for molecular properties, the generalized active space (GAS) technique is applied. We have considered 19 electrons as active, and imposed a 5$E_h$ (with $E_h$ referring to Hartree, the unit of energy in atomic units) virtual cut-off. In GAS model, an active space is divided into three subspaces: filled paired, unpaired, and virtuals named as GAS1, GAS2, and GAS3, respectively. The number of orbitals in each GAS and number of determinants with the active space are shown in Table~\ref{tab:table2}. The table also provides data on our choice of partitioning the occupied orbitals among GAS1 and GAS2, along with the PDMs obtained with that partitioning scheme. From our DZ results of the PDM, we choose the partitioning of GAS1 and GAS2 with 6 and 4 orbitals respectively for our TZ and QZ results, in view of the trade-off between the number of resulting determinants and precision in PDM. \\
\begin{table}[htbp]
\caption{\label{tab:table2}
Generalized active space model for the configuration interaction wave function of TlF$^+$ system with 5$E_h$ virtual cut-off energy for different basis sets. Within each basis, we have considered different choices for the partitioning of orbitals between GAS1 and GAS2, along with the corresponding value of PDM (given in Debye, abbreviated as D hereafter). }
\begin{ruledtabular}
\begin{tabular}{cccccccc}
 Basis set& GAS1& GAS2 & GAS3 & Number of & PDM (D)\\
 &&& &determinants&\\
\hline
DZ & 9 & 1 & 37 & 444602 & 2.21\\
  & 6 & 4 & 37 & 1249500 & 1.79 \\ 
  & 3 & 7 & 37 & 1519600 & 1.79\\
TZ & 9 & 1 & 56 & 1017642 & 2.28\\
  & 6 & 4 & 56 & 2870010 & 1.76\\
QZ & 9 & 1 &  88 & 2511530 & 2.29\\
   &6  & 4 & 88 & 7100730 & 1.83\\
\end{tabular}
\end{ruledtabular}
\end{table}

For RCC calculations, we have performed all-electron calculations by including all virtual orbitals at DZ and TZ levels. However, a 100 $E_h$ cut-off is imposed on virtuals to perform QZ calculation to make the computation manageable. The wave function itself is found by working in the RCCSD approximation, whereas only linear terms in $T$ are included to evaluate the expectation values of the molecular properties. This linear expectation value approximation is abbreviated as LERCCSD henceforth in this work. 
\section{Results and Discussion}
We now discuss our calculated results for  $E_{eff}$ and PDM (rounded off to the second decimal place) of TlF$^+$, using RCCSD and KRCISD approaches in different basis sets. The results are presented in Table~\ref{tab:table3}. Comparing results obtained from both the many-body methods enables us to understand the correlation effects captured by the two approaches. \\

We observe that the effective electric field is over 148 GV/cm. This is the largest known value of $E_{eff}$ among non-superheavy diatomic molecules. For comparison, $E_{eff}$ is about 115 GV/cm for HgF, while that of ThO is about 80 GV/cm. We have provided the effective electric fields of those molecules that provide the best upper limits (ThO, YbF, and HfF$^+$), and also other promising eEDM candidate molecules in Table~\ref{tab:table4}. We reiterate that a very large value of $E_{eff}$ translates to a significant enhancement of the expected sensitivity of an eEDM experiment. We now move to the electron correlation effects in $E_{eff}$ of TlF$^+$. The correlation effects captured by KRCISD contribute to about 2.5 percent to the quantity, while RCCSD changes the effective electric field by about 9 percent (in the QZ basis). We also observe that as we go from DZ to QZ basis, the effective electric field increases in the KRCI results, while it is found that the value of the property decreases when we use the RCCSD method. However, the net change as we go from DZ to QZ bases is less than 3 percent in the case of KRCI as well as RCCSD approaches, and the net change between the two methods with the QZ basis is less than 7 percent.  \\

On the other hand, we observe from Table~\ref{tab:table3} that the PDM of TlF$^+$ is very sensitive to the choice of many-body theory, with the KRCISD value of the quantity being almost one and a half times that of the RCCSD value. With the QZ basis sets, while KRCISD predicts 1.83 D, RCCSD approach yields a value of 1.26 D. From a many-body theoretic point of view, CCSD captures more correlation effects than CISD. However, in addition to this consideration, we also increased the number of active occupied spin-orbitals in our KRCISD computations. We note that our KRCISD calculations were carried out with 19 active electrons and a 5 $E_h$ virtuals cut-off. In order to understand the effect of inclusion of more electrons and virtual orbitals on the PDM, we performed two additional calculations at DZ level by considering (1) a virtual cut-off of 10$E_h$ with 19 active electrons (and the partitioning of occupied orbitals chosen as described in Section~\ref{sec:level3}), and found that the PDM changes by about 1 percent, and (2) 25 electrons with a virtual cut-off of 5 $E_h$ (with the partitioning again being the same as that from Section~\ref{sec:level3}), to find that the PDM changes by 0.6 percent. Increasing the number of active electrons beyond 25 would be accompanied by steep computational cost, hence we stop at that point. We expect that the inclusion of more electrons may lead to a slight lowering of PDM, but the extent to which it would finally agree with its RCCSD counterpart, would depend on the importance of correlation effects that are not captured by CISD but are present in CCSD. Lastly, we note that to enable comparison of the PDM of TlF$^+$ with other molecular ions listed in Table~\ref{tab:table4}, we have presented its PDM with the origin chosen to be the centre of mass, while in the rest of the manuscript, we specify the PDM with Tl chosen as the origin. \\

\begin{table}[htbp]
\caption{\label{tab:table3}
Calculated values of $\mu$ and $E_{eff}$ of TlF$^+$ at DF, KRCISD and RCCSD level of theory using different basis sets.}
\begin{ruledtabular}
\begin{tabular}{cccccccc}
 Basis set& $\mu$ (D)& $E_{eff}$ (GV/cm)\\
 \hline
 &\multicolumn{2}{c}{DF}\\
 \hline
 DZ & 2.75 & 140.49  \\
 TZ & 2.80 & 147.69 \\
 QZ & 2.79 & 148.74 \\
 \hline
 &\multicolumn{2}{c}{KRCISD}\\
\hline
DZ & 1.79 & 148.65 \\
TZ & 1.76 & 150.62 \\
QZ & 1.83 & 152.52 \\
\hline
&\multicolumn{2}{c}{LERCCSD}\\
\hline
DZ & 1.18 & 167.83 \\
TZ & 1.18 & 164.88  \\
QZ & 1.26 & 163.31 \\
\end{tabular}
\end{ruledtabular}
\end{table}

\begin{table}[htbp]
\caption{\label{tab:table4}
Comparison of computed PDM (in molecular frame) and $E_{eff}$ for TlF$^+$ using LERCCSD/QZ method with the other important non-superheavy molecules. }
\begin{ruledtabular}
\begin{tabular}{cccccccc}
 Molecule & $\mu$ (D) & $E_{eff}$ (GV/cm) & Ref.\\
 \hline
TlF$^+$ & 2.06 & 163.31  & This work\\
ThO & 4.24 &79.9 & \cite{Skripnikov_ThO_2016} \\
    & 4.41 & 75.2  & \cite{Malika_ThO_2016, Fleig_ThO_JMS2014}\\
HfF$^+$ & $-$ & 22.5  & \cite{Skripnikov_HfF_2017}\\
     & 3.81 & $-$ &  \cite{Leanhardt_2011}\\
ThF$^+$ & 4.03 & $-$  & \cite{Denis_2015}\\
PtH$^+$ & $-$ & 73  & \cite{Meyer_2006}\\
YbF & 3.60 & 23.1 & \cite{Abe_YbF_2014}\\
HgF & 3.45  & 113.77  & \cite{Prasannaa_sym}\\
HgCl & 3.45 & 110.94 &\cite{Prasannaa_sym}\\
HgBr & 2.94 &  107.42 & \cite{Prasannaa_sym}\\
HgI & 2.01 &  107.38 &  \cite{Prasannaa_sym}\\
RaH & 4.44 &  80.31 &\cite{Fazil_RaH_2019}\\
RaF & 3.85 & 52.5  &\cite{Sasmal_RaF_2016}\\
HgH & 0.27 & 123.2 &  \cite{Sasmal_HgH_2016}\\
    & 0.15 & 118.5 &  \cite{Sunaga_HgH_2017}\\
\end{tabular}
\end{ruledtabular}
\end{table} 

\begin{table}[ht]
\caption{\label{tab:table5}
Contributions of individual terms in the LERCCSD method to PDM (in D) using different basis sets. The nuclear contribution to the PDM is 84.30 D. }
\begin{ruledtabular}
\begin{tabular}{cccccccc}
 Basis & \\
 \hline
       &DF & $OT_1$ & $T_1^{\dag}O$\\
\hline
 DZ & 87.04 & -0.70 & -0.70 \\
 TZ & 87.09 & -0.69 & -0.69 \\
 QZ & 87.09 & -0.65 & -0.65\\
 \hline
   & $OT_2$ & $T_1^+OT_1$ & $T_1^+OT_2$\\
\hline
DZ & 0.0 & -0.15 & 0.11 \\
TZ & 0.0 & -0.25 & 0.12 \\
QZ & 0.0 & -0.23 & 0.11 \\
 \hline
   & $T_2^{\dag}O$ & $T_2^{\dag}OT_1$ & $T_2^{\dag}OT_2$\\
\hline
DZ & 0.0 & 0.11 & -0.22 \\
TZ & 0.0 & 0.12 & -0.22 \\
QZ & 0.0 & 0.11 & -0.22\\
\end{tabular}
\begin{flushleft}
*The operator $'O'$ in the table~\ref{tab:table5} is the electronic PDM operator i.e. second term in eq.~\ref{DM}.
\end{flushleft}
\end{ruledtabular}
\end{table}

\begin{table}[ht]
\caption{\label{tab:table6}
Contributions of individual terms from the LERCCSD method to $E_{eff}$ (in GV/cm) with different basis sets. }
\begin{ruledtabular}
\begin{tabular}{cccccccc}
 Basis & \\
\hline
   & DF & $OT_1$ & $T_1^{\dag}O$\\
\hline
DZ & 140.49 & 18.23 & 18.23 \\
TZ & 147.69 & 15.40 & 15.40 \\
QZ & 148.73 & 13.87 & 13.87\\
\hline
   & $OT_2$ & $T_1^{\dag}OT_1$ & $T_1^{\dag}OT_2$\\
\hline
DZ & 0.0 & -1.31 & 0.23 \\
TZ & 0.0 & -4.21 & -0.10 \\
QZ & 0.0 & -3.90 & -0.13 \\
\hline
   & $T_2^{\dag}O$ & $T_2^{\dag}OT_1$ & $T_2^{\dag}OT_2$\\
\hline
DZ & 0.0 & 0.23 & -8.27\\
TZ & 0.0 & -0.10 &  -9.19\\
QZ & 0.0 & -0.13 & -8.99\\
\end{tabular}
\begin{flushleft}
*The operator $'O'$ in Table~\ref{tab:table6} is the EDM operator, $H_{EDM}$.
\end{flushleft}
\end{ruledtabular}
\end{table}

In the LERCCSD approximation, the expectation value of an operator $O$ ($O$ could be $H_{EDM}$ or the PDM operator) is given by $\langle \Phi_0 \arrowvert (1+T_1+T_2)^\dag O (1+T_1+T_2) \arrowvert \Phi_0 \rangle_c$. The subscript, `c',  refers to the fact that each of the terms in the resulting expression are fully connected~\cite{Bartlett_2009}. Further details of the implementation can be found in Ref.~\cite{Abe_YbF_2014}. In tables ~\ref{tab:table5} and~\ref{tab:table6}, we present these results for $E_{eff}$ and PDM, to identify the dominant correlation contributions to these properties. It can be seen from these tables that $OT_1$ is the dominant term, followed by $T_2^\dag O T_2$ and then $T_1^\dag O T_1$. The contributions from $T_1^\dag O T_2$ and its hermitian conjugate terms are negligible for $E_{eff}$. We briefly comment on the nature of cancellation among the individual terms in, for example, $E_{eff}$, across basis sets. We see that as we move from DZ to QZ, the DF contribution increases while the $OT_1$ (and its hermitian conjugate) terms decrease, and thus the net change in DF + $OT_1$ + $T_1^\dag O$ is not significant. Note that among the individual terms that are opposite in sign to DF and $OT_1$ terms, $T_1^\dag O T_2$ and its hermitian conjugate are not significant (although there is a sign flip from DZ to TZ), $T_2^\dag OT_2$ changes little from DZ through QZ basis, and $T_1^\dag OT_1$ increases, such that the overall effective electric field changes within 3 percent between DZ and QZ bases. \\

We now comment on the potential sources of error in our RCCSD calculations. We anticipate that the possible error from not going to a higher quality basis is less than the difference between results obtained from QZ and TZ bases. We find that the PDM changes by about 6.5 percent, while $E_{eff}$ changes by less than 1 percent. Next, we examine the possible error from the LERCCSD approximation. We consider an earlier work that goes beyond the linear expectation value approximation~\cite{Prasannaa_sym} by considering several non-linear terms in $T$ in the expectation value. The work does not consider TlF$^+$, but considers a whole host of single valence systems. We find that while the effective electric field changes by at most 2.5 percent among the systems considered in that work (and hence assume that $E_{eff}$ for TlF$^+$ may change by a comparable amount), the PDM could change by as much as 26 percent. At this point, we conclude that the possible error in PDM can be large, and hence defer further analysis for a future study. We henceforth continue with only the error estimate for $E_{eff}$. We recall that the error due to our choice of an offset bond length is about 10 percent. We assume that neglecting higher order excitations such as triples etc would not change the effective electric field by over 5 percent. Lastly, we do not expect that increasing the cut-off on high-lying virtuals beyond 100 $E_h$ would make any difference to $E_{eff}$. We conclude that the expected error in the calculated value of $E_{eff}$ is at most 18 percent.\\ 
\section{Summary}
In summary, based on our relativistic many-body theoretic analyses, we conclude that TlF$^+$ could be a promising molecular candidate for future electron electric dipole moment searches. TlF$^+$ possesses a large effective electric field of 163.31 GV/cm, which is by far the largest known value for the property for a non-superheavy molecule. We also study the observed electron correlation trends in the effective electric field as well as the molecular electric dipole moment of the molecule, across basis sets and using KRCI and the RCCSD methods. We find that while correlation effects are crucial in determining the value of PDM, it is not as critical for the effective electric field. In order to exploit the large effective electric field in TlF$^+$, a viable experimental method needs to be investigated. \\ 
\begin{acknowledgments}
We acknowledge Prof. Amar Vutha and Prof. Andrew Jayich for discussions on experimental aspects. We are also thankful to Prof. Debashis Mukherjee for fruitful discussions on one of the theoretical frameworks employed in this work. All the calculations are performed on Rudra cluster, Sankhyasutra Labs (Bangalore, India), and computational facility available in Japan. This work was supported by JSPS KAKENHI Grant Numbers JP21H01864 and JP20H01929. 
\end{acknowledgments}
\nocite{*}

\bibliography{TlF+}

\end{document}